\documentclass[conference]{IEEEtran}
\IEEEoverridecommandlockouts

\usepackage{cite}
\usepackage{amsmath,amssymb,amsfonts}
\usepackage{algorithmic}
\usepackage{graphicx}
\usepackage{textcomp}
\usepackage{xcolor}

\usepackage{float}
\usepackage{array}
\usepackage{multirow}
\usepackage{booktabs}
\usepackage[table]{xcolor}
\usepackage[hidelinks]{hyperref}
\usepackage{algorithm}

\definecolor{stsoBlue}{RGB}{230,240,255}

\def\BibTeX{{\rm B\kern-.05em{\sc i\kern-.025em b}\kern-.08em
    T\kern-.1667em\lower.7ex\hbox{E}\kern-.125emX}}
\begin{document}

\title{Spatio-Temporal Wireless-Optical Planning for Multi-UAV Networks\\

\author{
\IEEEauthorblockN{
Binglei Wang\textsuperscript{1,2},
Huiru Ao\textsuperscript{2},
Fan Yang\textsuperscript{1,2},
Zhenjie Zhou\textsuperscript{1,2},
Zhonghua Peng\textsuperscript{2},
Jialong Li\textsuperscript{2,*}
}
\IEEEauthorblockA{
\textsuperscript{1}Southern University of Science and Technology\\
\textsuperscript{2}Faculty of Computer Science and Artificial Intelligence, 
Shenzhen University of Advanced Technology\\
}
}

}


\maketitle

\begin{abstract}
Multi-unmanned aerial vehicle (UAV) networks in urban low-altitude environments couple UAV mobility, wireless access, and optical backhaul resources. Existing path-planning methods optimize flight distance or wireless signal quality, but can still concentrate traffic on shared optical backhaul links. We present Spatio-Temporal Wireless-Optical (STWO) planner, a backhaul-aware path-planning algorithm that jointly considers flight distance, wireless link quality, and time-varying optical-link offered-load ratio. STWO updates backhaul occupancy during sequential multi-UAV planning, allowing later UAVs to avoid congested optical paths while maintaining wireless connectivity. Experiments show that STWO reduces peak optical-link offered-load ratio by up to 56.4\% and congestion ratio by up to 72.6\% under dense UAV deployment, demonstrating the importance of wireless-optical awareness for reliable multi-UAV transmission.
\end{abstract}


\section{Introduction}
With the development of the low-altitude economy, smart-city construction, and 5G/6G wireless networks, multiple unmanned aerial vehicles (UAVs) are increasingly used in urban inspection, security monitoring, emergency communications, traffic surveillance, and other mission-oriented scenarios~\cite{ullah2020cognition,geraci2022will}. In these applications, UAVs are no longer isolated platforms, but aerial user equipment and cooperative sensing nodes that continuously interact with ground cellular networks~\cite{banafaa2024comprehensive}. Recent work shows that uplink congestion affects low latency mobile uploading~\cite{xu2024congestion}. Their mobility determines not only flight distance and obstacle avoidance, but also serving base-station selection and end-to-end transmission performance. As UAVs continuously upload video, images, and sensing data during flight, path planning must account for communication reliability beyond the air-to-ground wireless link.

Existing communication-aware path planning methods focus mainly on wireless coverage quality, link reliability, trajectory optimization, and handover reduction~\cite{zhang2020radio,guo2021uav,du2025handover}. Radio-map-based trajectory planning and learning-based path optimization have been shown to improve the communication quality of cellular-connected UAVs~\cite{zhang2020radio,li2022path}. However, these methods largely optimize the access link while overlooking the optical backhaul after UAV traffic enters base stations. In urban low-altitude environments, high-rise blockage may weaken wireless signals, while multiple UAVs may also be attracted to strong-signal base stations. This can concentrate traffic on shared aggregation node--core network (AGG--Core) optical backhaul links, creating transport-side bottlenecks even when the wireless access links appear reliable. Recent studies on metro-access convergence, optical time-slice switching, and multidimensional resource allocation indicate that transport-side resources can become bottlenecks under dynamic traffic demand~\cite{li2019flexible,li2017balancing,li2022leveraging,li2018flexible,li2020towards}. Optical access and x-haul studies further show that access, aggregation, and core transport resources should be jointly considered in 5G/6G networks~\cite{ranaweera20224g,ranaweera2023design}. These observations call for path planning that jointly balances flight distance, wireless quality, and optical-backhaul load.

To address this issue, we present Spatio-Temporal Wireless-Optical (STWO) planner, a backhaul-aware path-planning algorithm for multi-UAV networks. STWO incorporates flight distance, wireless link quality, and time-varying optical-link offered-load ratio into a unified path-search cost. During planning, STWO evaluates candidate serving base stations using a joint wireless-optical cost and maps UAV traffic to the corresponding base station--aggregation node--core network (BS--AGG--Core) optical backhaul path. For multi-UAV scenarios, STWO adopts a sequential planning mechanism: after each UAV path is planned, the optical-link occupancy is updated over time, allowing subsequent UAVs to perceive existing backhaul load and avoid congested optical paths. Compared with Vanilla A* and Signal-Aware A*, STWO reduces peak optical-link offered-load ratio by up to 56.4\% and congestion ratio by up to 72.6\% under dense UAV deployment, showing the benefit of joint wireless-optical awareness for reliable multi-UAV transmission.

The rest of this paper is organized as follows. 
Section~\ref{sec:model} models the coupling among UAV paths, wireless access, and optical backhaul resources. 
Section~\ref{sec:algorithm} presents the STWO algorithm and its path-search formulation. 
Section~\ref{sec:results} evaluates STWO under varying UAV densities, traffic rates, and backhaul capacities. 
Section~\ref{sec:conclusion} concludes the paper and discusses future directions.

\section{Coupled Wireless-Optical Model}\label{sec:model}

\begin{figure*}[t]
\vspace{-1.2em}
\centering
\includegraphics[width=0.98\textwidth]{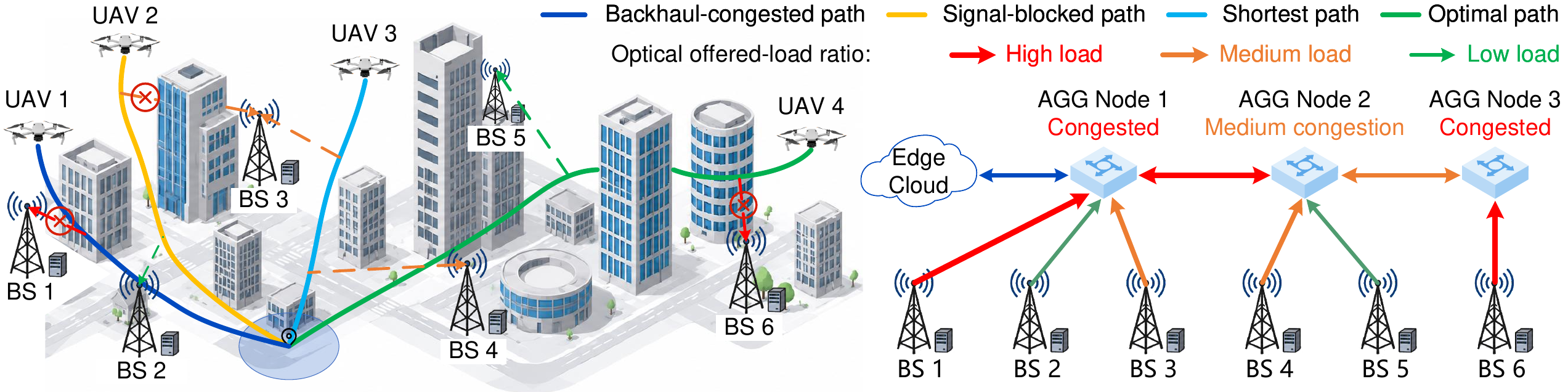}
\caption{Wireless-optical coupled model for urban low-altitude multi-UAV path planning. UAV paths determine serving base-station selection, which further maps traffic onto AGG--Core optical backhaul links.}
\label{fig:system_model}
\vspace{-1.2em}
\end{figure*}

As shown in Fig.~\ref{fig:system_model}, we consider an urban low-altitude multi-UAV path-planning scenario under cellular coverage. A UAV path affects not only flight distance and blockage conditions, but also the sequence of serving base stations along the route~\cite{zhang2020radio,guo2021uav}. Once UAV traffic enters the cellular network, the selected base station further determines the corresponding BS--AGG--Core network optical backhaul path. Therefore, UAV mobility, wireless access, and optical backhaul load are inherently coupled. The BS--AGG--Core abstraction in Fig.~\ref{fig:system_model} captures this mapping between UAV access decisions and transport-network resource usage, which is consistent with 5G/6G transport and optical x-haul network design where access, aggregation, and core resources are jointly constrained by capacity and latency requirements~\cite{li2019flexible,ranaweera20224g,ranaweera2023design}.

This coupling creates different outcomes for different path-planning objectives. If path planning considers only flight distance, UAVs may traverse areas blocked by high-rise buildings, causing wireless signal attenuation and link reliability degradation~\cite{zhang2020radio,du2025handover}. If it considers only wireless signal quality, multiple UAVs may be attracted to a small number of strongly covered base stations, concentrating traffic on shared optical backhaul links. For example, in Fig.~\ref{fig:system_model}, traffic accessing BS~6 is aggregated to the already congested AGG Node~3, whereas accessing BS~5 can map the traffic to a less loaded backhaul path. Existing studies on metro-access convergence, optical transport slicing, and backhaul-aware association also indicate that transport-side capacity constraints may become bottlenecks even when the wireless access link remains acceptable~\cite{li2019flexible,li2020end,nafees2023backhaul}.

These observations show that urban low-altitude multi-UAV path planning is a cross-layer planning problem across flight paths, wireless access, and optical backhaul resources. A desirable planner should avoid wireless-blocked regions while also preventing multiple UAVs from overloading the same AGG--Core optical links. This motivates a wireless-optical planning that jointly accounts for flight distance, wireless connection quality, and time-varying optical backhaul offered-load ratio.

\section{STWO Algorithm Design}\label{sec:algorithm}

This section presents the Spatio-Temporal Wireless-Optical (STWO) planner, a
heuristic path-search algorithm for multi-UAV networks. STWO searches over
space-time states and jointly accounts for UAV position, arrival time, wireless
link quality, serving base-station selection, and optical-link offered-load ratio.
Unlike Vanilla A$^*$, which minimizes flight distance, and Signal-Aware A$^*$,
which mainly follows wireless quality, STWO incorporates time-varying
optical-link offered-load ratio into the path-search process.

STWO extends the conventional two-dimensional spatial node into a space-time
state:
\begin{equation}
s=(x,y,t),
\label{eq:space_time_node}
\end{equation}
where $(x,y)$ denotes the grid position of the UAV, and $t$ denotes the time
step at which the UAV reaches that position. Let $\mathcal{U}$ denote the UAV
set, $\mathcal{B}$ the base-station set, and $\mathcal{L}$ the optical backhaul
link set. For each base station $b$, we use $\mathcal{P}_b\subseteq\mathcal{L}$
to denote its BS--AGG--Core optical backhaul path. When a UAV accesses $b$, its
traffic is carried by the links in $\mathcal{P}_b$.

The optical-link offered-load ratio of link $l$ at time $t$ is defined as
\begin{equation}
U_l(t)=\frac{\mathrm{Load}_l(t)}{\mathrm{Capacity}_l},
\label{eq:link_offered_load}
\end{equation}
where $\mathrm{Load}_l(t)$ and $\mathrm{Capacity}_l$ denote the offered traffic load
and capacity of link $l$, respectively. Here, $U_l(t)$ represents the traffic
demand normalized by the link capacity. A value larger than 1 indicates
demand oversubscription, rather than physical link utilization above 100\%.

The received signal strength from base station $b$ at position $p$ is modeled
as~\cite{zhang2020radio,chen2024optimal,li2025joint}
\begin{equation}
\mathrm{RSRP}(p,b)=P_{0}-20\log_{10} d(p,b)-L_{\mathrm{block}},
\label{eq:rsrp}
\end{equation}
where $P_0$ is a calibrated reference-power offset, $d(p,b)$ is the distance between
position $p$ and base station $b$, and $L_{\mathrm{block}}$ is the blockage loss. This simplified model provides a calibrated dBm-scale RSRP value for comparing
wireless link quality across different UAV paths, rather than a site-specific
field-measured received power.

For each candidate base station, STWO defines wireless and optical penalties.
The wireless penalty measures the RSRP deficit and is defined as~\cite{gao2021cellular}
\begin{equation}
C_{\mathrm{sig}}(p,b)=
\max\left(0,\mathrm{RSRP}_{\mathrm{tar}}-\mathrm{RSRP}(p,b)\right),
\label{eq:signal_cost}
\end{equation}
where $\mathrm{RSRP}_{\mathrm{tar}}$ is the target signal level. The maximum
optical-link offered-load ratio on the backhaul path of $b$ is
\begin{equation}
U_{\max}(b,t)=\max_{l\in\mathcal{P}_b} U_l(t).
\label{eq:max_utilization}
\end{equation}
The optical load penalty is defined as
\begin{equation}
C_{\mathrm{opt}}(b,t)=
U_{\max}(b,t)+k\max\left(0,U_{\max}(b,t)-\theta\right)^2,
\label{eq:optical_cost}
\end{equation}
where $\theta$ is the congestion-warning threshold and $k$ controls the penalty
for exceeding the threshold.

At each space-time state, STWO selects the serving base station by balancing
wireless quality and optical offered-load ratio:
\begin{equation}
b^*(p,t)=\arg\min_{b\in\mathcal{B}}
\left(\beta C_{\mathrm{sig}}(p,b)+\gamma C_{\mathrm{opt}}(b,t)\right).
\label{eq:serving_bs}
\end{equation}
The transition cost is then defined as
\begin{equation}
C=\alpha C_{\mathrm{dist}}
+\beta C_{\mathrm{sig}}(p,b^*(p,t))
+\gamma C_{\mathrm{opt}}(b^*(p,t),t),
\label{eq:total_cost}
\end{equation}
where $C_{\mathrm{dist}}$ is the flight-distance cost, and $\alpha$, $\beta$, and
$\gamma$ are the weights for distance, wireless quality, and optical offered-load ratio, respectively. STWO follows the A$^*$ evaluation function
\begin{equation}
f(n)=g(n)+h(n),
\label{eq:astar_function}
\end{equation}
where $g(n)$ is the accumulated path cost from the source to node $n$, and
$h(n)$ is the heuristic distance estimate from node $n$ to the destination.

\begin{figure*}[t]
\centering
\includegraphics[width=1.0\textwidth]{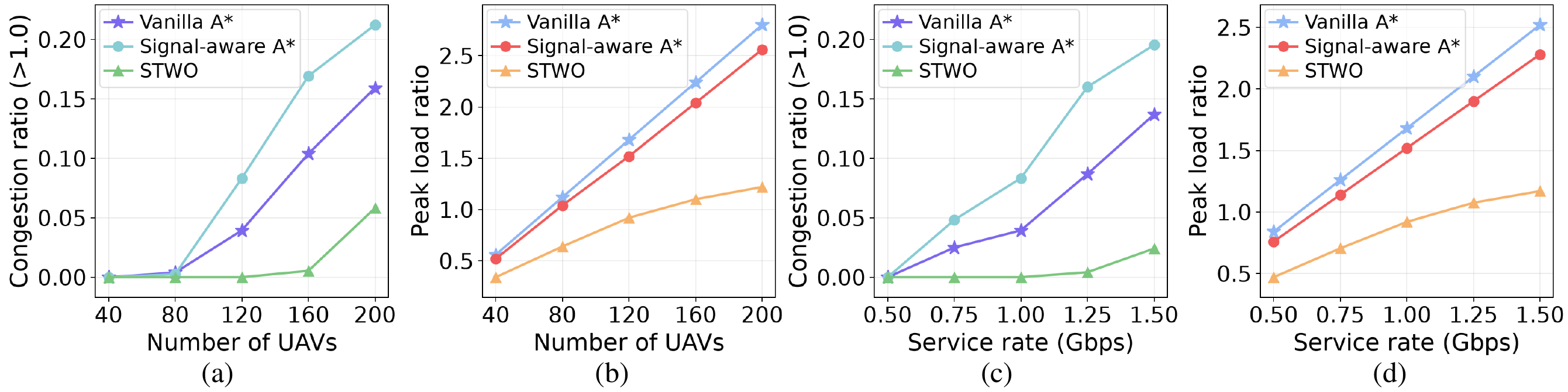}
\vspace{-1.5em}
\caption{Network performance under different UAV numbers and traffic rates: 
(a) congestion ratio versus UAV number; 
(b) peak optical-link offered-load ratio versus UAV number; 
(c) congestion ratio versus traffic rate; 
(d) peak optical-link offered-load ratio versus traffic rate.}
\label{fig:uav_rate_results}
\vspace{-1.2em}
\end{figure*}

\begin{algorithm}[t]
\caption{STWO planner}
\label{alg:stwo}
\footnotesize
\begin{algorithmic}[1]
\REQUIRE UAV set $\mathcal{U}$, base-station set $\mathcal{B}$, optical links $\mathcal{L}$, backhaul paths $\{\mathcal{P}_b\}$, UAV traffic rates $\{r_u\}$
\ENSURE UAV paths $\{\pi_u\}$ and serving base-station sequences $\{b_u(t)\}$
\STATE Initialize $\mathrm{Load}_l(t)\gets 0$ for all $l\in\mathcal{L}$ and time step $t$
\FOR{each UAV $u\in\mathcal{U}$}
    \STATE Initialize the open set with the source state of UAV $u$
    \WHILE{the open set is not empty}
        \STATE Select the state $n=(p,t)$ with the minimum $f(n)$
        \IF{$n$ reaches the destination}
            \STATE Recover $\pi_u$ and $b_u(t)$; \textbf{break}
        \ENDIF
        \FOR{each feasible next state $(p',t+1)$}
            \STATE Select $b^*(p',t+1)$ by Eq.~\eqref{eq:serving_bs}
            \STATE Compute transition cost by Eq.~\eqref{eq:total_cost}
            \STATE Update $g(\cdot)$, $f(\cdot)$, and parent state
        \ENDFOR
    \ENDWHILE
    \STATE Update $\mathrm{Load}_l(t)$ along $\mathcal{P}_{b_u(t)}$ using rate $r_u$ for all states on $\pi_u$
\ENDFOR
\STATE \textbf{return} $\{\pi_u\}$ and $\{b_u(t)\}$
\end{algorithmic}
\end{algorithm}

As shown in Alg.~\ref{alg:stwo}, STWO plans UAV paths sequentially. For each
candidate space-time state, STWO selects the serving base station using the
wireless-optical cost in Eq.~\eqref{eq:serving_bs} and evaluates the transition
cost by Eq.~\eqref{eq:total_cost}. After one UAV path is obtained, STWO updates
the time-indexed optical-link load along the selected backhaul paths. Subsequent
UAVs therefore perceive existing backhaul occupancy and avoid congestion conflicts
within the same time period. As a result, STWO guides UAVs away from high-load
optical paths while maintaining wireless connectivity, thereby balancing flight
distance, wireless access, and optical backhaul offered load.

\section{Experimental Results and Analysis}\label{sec:results}

This section evaluates STWO under varying UAV densities, traffic rates, and
backhaul capacities. We compare it with distance-based and signal-aware baselines
to examine whether STWO reduces optical-link congestion while preserving wireless
connectivity and moderate path length.

\subsection{Experimental Setup}

We evaluate STWO in a grid-based urban low-altitude scenario. The simulated area is
$1000~\mathrm{m}\times1000~\mathrm{m}$ and is discretized into $100\times100$ grid
cells with a grid resolution of $10~\mathrm{m}$. UAVs fly at an altitude of
$100~\mathrm{m}$ and share the same destination at $(900,900)$. The scenario contains
three buildings, five base stations, and three aggregation nodes. Each base station
is mapped to one AGG--Core optical backhaul path. Unless otherwise specified, the
BS--AGG link capacity is $50~\mathrm{Gbps}$, the AGG--Core link capacity is
$100~\mathrm{Gbps}$, the number of UAVs is 120, and the per-UAV traffic rate is
$1000~\mathrm{Mbps}$. The background load is set to zero, so the optical-link load is
generated only by UAV traffic.

Table~\ref{tab:sim_params} summarizes the key simulation parameters. We include the
parameters that directly affect wireless quality, optical backhaul load, and STWO
path-search cost.

\begin{table}[t]
\caption{Key Simulation Parameters}
\label{tab:sim_params}
\centering
\footnotesize
\setlength{\tabcolsep}{3pt}
\renewcommand{\arraystretch}{1.12}
\begin{tabular}{l|l}
\toprule
\textbf{Parameter} & \textbf{Setting} \\
\midrule
Map size & $1000~\mathrm{m}\times1000~\mathrm{m}$ \\
Grid resolution & $10~\mathrm{m}$, $100\times100$ cells \\
UAV altitude & $100~\mathrm{m}$ \\
Destination & $(900,900)$ \\
Buildings / BSs / AGGs & 3 / 5 / 3 \\
Default UAV number & 120 \\
UAV number sweep & 40, 80, 120, 160, 200 \\
Default per-UAV rate & $1000~\mathrm{Mbps}$ \\
BS--AGG capacity & $50~\mathrm{Gbps}$ \\
Default AGG--Core capacity & $100~\mathrm{Gbps}$ \\
Traffic-rate sweep & 500, 750, 1000, 1250, 1500 Mbps \\
AGG--Core capacity sweep & 50, 60, 70, 80, 90, 100 Gbps \\
Calibrated $P_0$ & $-35~\mathrm{dBm}$ \\
$L_{\mathrm{block}}$ & $25~\mathrm{dB}$ \\
$\mathrm{RSRP}_{\mathrm{tar}}$ & $-76~\mathrm{dBm}$ \\
Cost weights $(\alpha,\beta,\gamma)$ & $(1.0, 2.5, 1800.0)$ \\
Optical penalty $(\theta,k)$ & $(0.8, 16.0)$ \\
Offered-load ratio & $U_l(t)=\mathrm{Load}_l(t)/\mathrm{Capacity}_l$ \\
Congestion threshold & $U_l(t)>1.0$ \\
\bottomrule
\end{tabular}
\vspace{-1.5em}
\end{table}

We compare STWO with Vanilla A$^*$, which minimizes flight distance, and Signal-Aware A$^*$, which considers wireless signal quality but ignores optical backhaul load. Sequential Optical-only A$^*$ is further used for ablation. The metrics include PeakLoad, congestion ratio, hotspot ratio, average RSRP, and average path length. PeakLoad denotes the peak optical-link offered-load ratio, i.e., the maximum demand-to-capacity ratio. Values above 100\% indicate demand oversubscription rather than physical utilization above 100\%. Cong ($>1.0$) and Hotspot ($>0.8$) count optical-link samples whose offered-load ratios exceed 1.0 and 0.8, respectively.

\subsection{Impact of UAV Density and Traffic Rate}

Figs.~\ref{fig:uav_rate_results}(a) and~(b) show the impact of UAV density. As the
number of UAVs increases from 40 to 200, Vanilla A$^*$ and Signal-Aware A$^*$
show rapidly increasing congestion and peak optical-link offered-load ratio.
Vanilla A$^*$ ignores wireless and backhaul conditions, while Signal-Aware A$^*$
may concentrate traffic on strongly covered base stations and shared backhaul
paths. In contrast, STWO considers time-varying optical backhaul load and keeps
traffic more balanced. At $N=200$, STWO reduces the peak offered-load ratio from
280.00\% and 256.00\% to 122.00\%, corresponding to reductions of 56.4\% and
52.3\%, respectively. It also reduces the congestion ratio from 15.92\% and
21.22\% to 5.82\%.

\begin{table*}[t]
\caption{Experimental Results Under Different Settings}
\label{tab:exp_results}
\centering
\footnotesize
\setlength{\tabcolsep}{4pt}
\renewcommand{\arraystretch}{1.15}
\begin{tabular}{c|l|ccc|cc}
\toprule
\multirow{2}{*}{\textbf{Experiment Setting}} 
& \multirow{2}{*}{\textbf{Method}} 
& \multicolumn{3}{c|}{\textbf{Congestion-related Metrics}} 
& \multicolumn{2}{c}{\textbf{Communication/Path Metrics}} \\
\cmidrule(lr){3-5} \cmidrule(lr){6-7}
& 
& \textbf{PeakLoad (\%)} 
& \textbf{Cong ($>1.0$) (\%)} 
& \textbf{Hotspot ($>0.8$) (\%)} 
& \textbf{AvgRSRP (dBm)} 
& \textbf{AvgLen (m)} \\
\midrule

\multirow{3}{*}{Scalability ($N=200$)} 
& Vanilla A$^*$      & 280.00 & 15.92 & 20.26 & -78.05 & 888.8 \\
& Signal-Aware A$^*$ & 256.00 & 21.22 & 23.83 & -75.30 & 1042.7 \\
& \cellcolor{stsoBlue}STWO 
& \cellcolor{stsoBlue}122.00 
& \cellcolor{stsoBlue}5.82 
& \cellcolor{stsoBlue}16.78 
& \cellcolor{stsoBlue}-77.21 
& \cellcolor{stsoBlue}1128.2 \\
\midrule

\multirow{3}{*}{Traffic rate ($1500$ Mbps)} 
& Vanilla A$^*$      & 252.00 & 13.68 & 18.95 & -78.05 & 889.1 \\
& Signal-Aware A$^*$ & 228.00 & 19.53 & 22.53 & -75.30 & 1042.9 \\
& \cellcolor{stsoBlue}STWO 
& \cellcolor{stsoBlue}117.00 
& \cellcolor{stsoBlue}2.41 
& \cellcolor{stsoBlue}13.28 
& \cellcolor{stsoBlue}-76.97 
& \cellcolor{stsoBlue}1100.1 \\
\midrule

\multirow{3}{*}{Backhaul capacity ($50$ Gbps)} 
& Vanilla A$^*$      & 168.00 & 11.18 & 19.74 & -78.05 & 889.1 \\
& Signal-Aware A$^*$ & 192.00 & 16.67 & 25.78 & -75.30 & 1042.9 \\
& \cellcolor{stsoBlue}STWO 
& \cellcolor{stsoBlue}108.00 
& \cellcolor{stsoBlue}1.14 
& \cellcolor{stsoBlue}5.98 
& \cellcolor{stsoBlue}-76.78 
& \cellcolor{stsoBlue}1029.1 \\
\midrule

\multirow{4}{*}{Ablation study} 
& Vanilla A$^*$                 & 336.00 & 20.13 & 37.63 & -78.05 & 889.1 \\
& Signal-Aware A$^*$            & 304.00 & 23.70 & 40.36 & -75.30 & 1042.9 \\
& Sequential Optical-only A$^*$ & 140.00 & 14.37 & 21.06 & -78.92 & 1161.6 \\
& \cellcolor{stsoBlue}STWO 
& \cellcolor{stsoBlue}140.00 
& \cellcolor{stsoBlue}15.29 
& \cellcolor{stsoBlue}20.79 
& \cellcolor{stsoBlue}-77.26 
& \cellcolor{stsoBlue}1154.8 \\
\bottomrule
\end{tabular}
\vspace{-1.2em}
\end{table*}

Figs.~\ref{fig:uav_rate_results}(c) and~(d) show the impact of traffic rate. As the
per-UAV rate increases from 0.5 Gbps to 1.5 Gbps, Vanilla A$^*$ and Signal-Aware
A$^*$ suffer from increasing congestion and peak offered-load ratio. STWO keeps
the congestion ratio much lower across the tested rate range and limits the peak
offered-load ratio to 117.00\% at 1.5 Gbps. Compared with Vanilla A$^*$ and
Signal-Aware A$^*$, this reduces the peak offered-load ratio by 53.6\% and
48.7\%, respectively. These results show that STWO is more robust to
bandwidth-intensive UAV traffic.

\vspace{-0.1in}
\subsection{Impact of Backhaul Capacity}

\begin{figure}[!t]
\centering
\includegraphics[width=1.0\columnwidth]{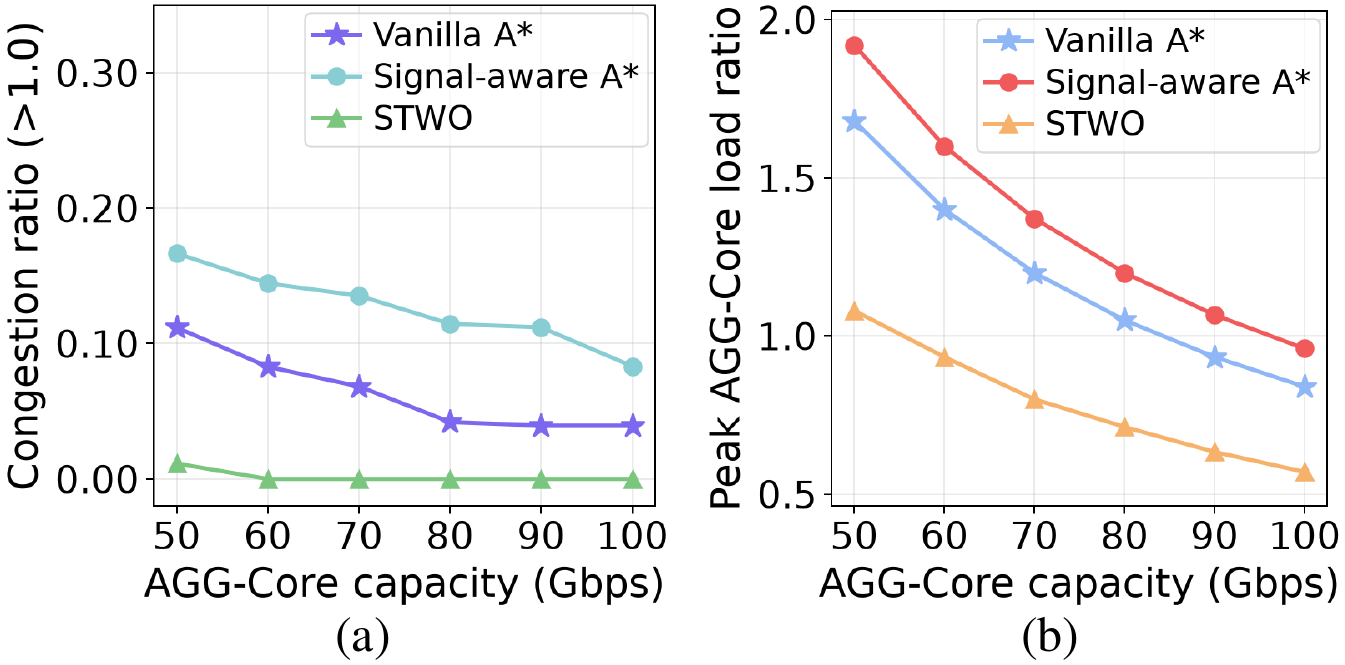}
\vspace{-1.5em}
\caption{Network performance under different AGG–Core optical-backhaul capacities:
(a) congestion ratio and (b) peak AGG-Core offered-load ratio.}
\label{fig:capacity_results}
\vspace{-1.2em}
\end{figure}

Figs.~\ref{fig:capacity_results}(a) and~(b) analyze the impact of AGG--Core
capacity. As the capacity increases from 50~Gbps to 100~Gbps, all methods
experience lower congestion ratios and lower peak AGG--Core offered-load ratios.
STWO consistently achieves the lowest peak AGG--Core offered-load ratio across
the entire capacity range. At 50~Gbps, STWO reduces the peak AGG--Core
offered-load ratio from 168.00\% and 192.00\% to 108.00\%, corresponding to
reductions of 35.7\% and 43.8\% compared with Vanilla A$^*$ and Signal-Aware
A$^*$, respectively. It also reduces the congestion ratio from 11.18\% and
16.67\% to 1.14\% under this setting.

\subsection{Hotspot Ratio and Ablation Study}

\begin{figure}[!t]
\centering
\includegraphics[width=1.0\columnwidth]{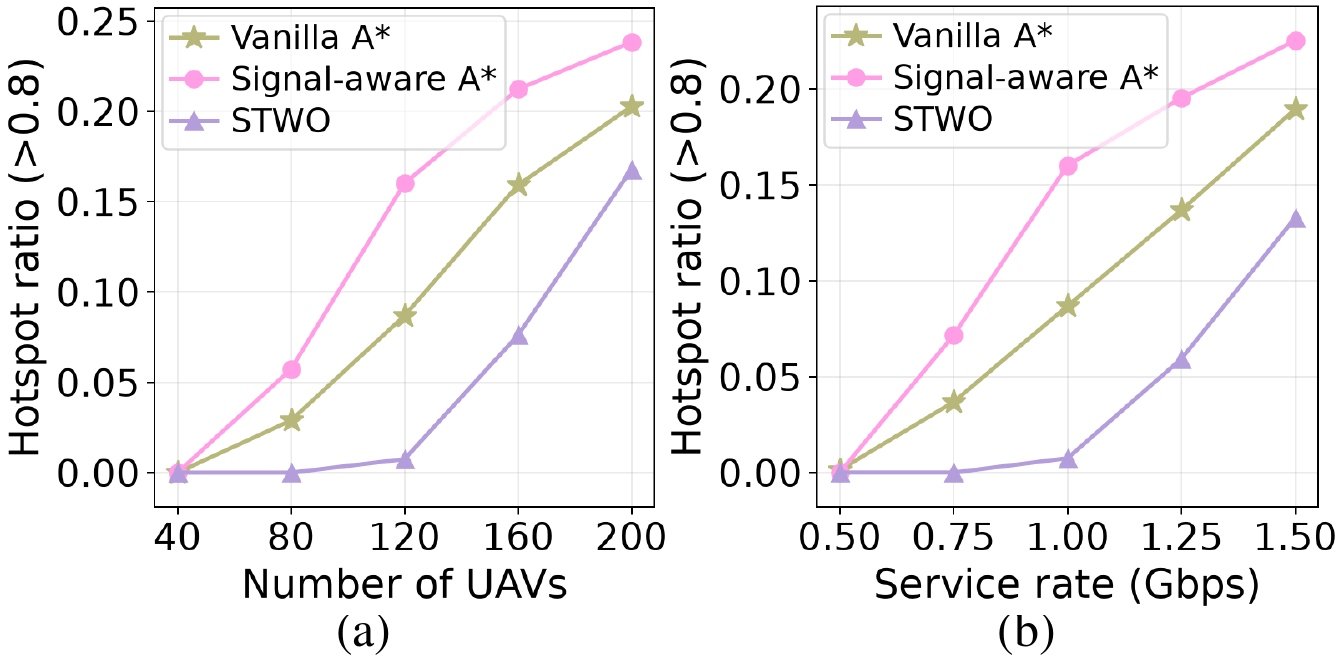}
\vspace{-1.5em}
\caption{Hotspot ratio under different load conditions:
(a) hotspot ratio versus UAV number and (b) hotspot ratio versus traffic rate.}
\label{fig:hot_congestion_duration}
\vspace{-1.2em}
\end{figure}

Figs.~\ref{fig:hot_congestion_duration}(a) and~(b) show the hotspot ratio,
where the optical-link offered-load ratio exceeds the warning threshold 0.8.
STWO consistently reduces the hotspot ratio compared with Vanilla A$^*$ and
Signal-Aware A$^*$ across all settings. For example, at $N=200$, STWO maintains
a lower hotspot ratio than both baselines. Under increasing traffic rates, STWO
also keeps fewer optical links in near-capacity operation, while the baselines
show higher hotspot ratios as traffic demand increases.

This result complements the Hotspot ($>0.8$) metric in Table~\ref{tab:exp_results}.
Hotspot ($>0.8$) measures near-capacity operation rather than true overload. In the
largest-scale case, STWO reduces the Hotspot ($>0.8$) ratio from 20.26\% and
23.83\% to 16.78\%, while also reducing the true congestion ratio from 15.92\% and
21.22\% to 5.82\%. This indicates that STWO not only suppresses severe overload, but
also reduces near-capacity hotspot operation.

The ablation results in Table~\ref{tab:exp_results} show the trade-off between
wireless quality and optical backhaul load. Signal-Aware A$^*$ achieves the best
AvgRSRP but leads to higher PeakLoad, while Sequential Optical-only A$^*$ reduces
optical load at the cost of degraded AvgRSRP. STWO achieves the same PeakLoad as
Sequential Optical-only A$^*$, improves AvgRSRP from $-78.92$ dBm to $-77.26$ dBm,
and only slightly increases the congestion ratio. This demonstrates that STWO
better balances wireless connectivity and optical backhaul load under the joint
objective.

\section{Conclusion}\label{sec:conclusion}

This paper presented STWO, a spatio-temporal wireless-optical planner for multi-UAV
transmission in urban low-altitude networks. By jointly considering flight distance,
wireless signal quality, and time-varying optical-link utilization, STWO coordinates
UAV path planning with access and backhaul conditions. Experiments show that STWO reduces the peak optical-link offered-load ratio and congestion ratio while maintaining competitive
wireless quality and path length. These results highlight the importance of joint
wireless and optical backhaul awareness for reliable multi-UAV transmission. Future
work will extend STWO to handover-aware trajectory optimization and backhaul resource
orchestration.

\bibliographystyle{IEEEtran}
\bibliography{references}

\end{document}